# A Model for Non-Linear Quantum Evolution based on Time Displaced Entanglement

T. C. Ralph,

Department of Physics, University of Queensland, Brisbane 4072, QLD, Australia

(Dated: October 5, 2018)

We discuss a model for non-linear quantum evolution based on the idea of time displaced entanglement, produced by taking one member of an entangled pair on a round trip at relativistic speeds, thus inducing a time-shift between the pair. We show that decoherence of the entangled pair is predicted. For non-maximal entanglement this then implies the ability to induce a non-unitary, non-linear quantum evolution. Although exhibiting unusual characteristics, we show that these evolutions cannot be dismissed on the basis of entropic or causal arguments.

## I. INTRODUCTION

A defining feature of separated quantum systems is the ability to entangle them, such that the correlations between the systems cannot be explained by local realistic theories [1]. A field of considerable recent interest is the study of entanglement in relativistic scenarios involving both inertial [2] and non-inertial frames [3]. The tools of quantum information science [4] can be used to investigate novel effects. The lack of a complete theoretical description of quantum processes in a relativistic framework leaves open the possibility that such studies might reveal fundamentally new phenomena.

Here we consider a simple model in which one member of an entangled pair suffers relativistic time dilation before being reunited with its pair. We refer to this as time displaced entanglement [5]. The model predicts that strong non-unitary and non-linear evolution can occur under certain conditions. Normally non-linear quantum evolutions would be discounted on the basis that they can lead to superluminal communication [6] and nonunitarity on thermodynamic grounds. However we show that our model is well behaved both causally and thermodynamically.

## II. TEMPORAL REPRESENTATION OF QUBITS

We consider abstract quantum two-level systems (qubits). Using field operators and dual-rail logic [7] we propose the following temporal mode representation of their logical states:

$$\begin{aligned}
|0\rangle_t &= \int_{t-\Delta}^{t+\Delta} dt' F(t'-t) \hat{a}_{t'}^\dagger |g\rangle_a |g\rangle_b \\
|1\rangle_t &= \int_{t-\Delta}^{t+\Delta} dt' F(t'-t) \hat{b}_{t'}^\dagger |g\rangle_a |g\rangle_b
\end{aligned} \quad (1)$$

where $\hat{a}_{t'}^\dagger$ and $\hat{b}_{t'}^\dagger$ are single time bosonic creation operators with the non-zero commutators $[\hat{a}_{t'}, \hat{a}_{t''}^\dagger] = [\hat{b}_{t'}, \hat{b}_{t''}^\dagger] = \delta(t'-t'')$ and corresponding ground-states $|g\rangle_a$ and $|g\rangle_b$. $F(t')$ is the temporal wave function of the qubit, defining its measurement and interaction bandwidths, and related by Fourier transform to the energy spectra of the qubit states [8]. We have assumed for convenience that $F(t') = 0$ for $t' \geq |\Delta|$ thus allowing us to bound the limits of the integrals [9]. Also for convenience we introduce an $nth$ clock-cycle time, $t_n$ such that $t_n - t_{n-1} > 2\Delta$. A measurement (or gate) at the $nth$ clock-cycle must be carried out between times $t_n - \Delta$ and $t_n + \Delta$ to definitely detect (or transform) the qubit state. The qubit states $|0\rangle_{t_n}$ and $|1\rangle_{t_n}$ form a complete basis for all the possible qubit states at the $nth$ clock-cycle. Quantum optical systems can be described via a similar formalism [10]. Time evolution of an arbitrary qubit can be represented by:

$$\begin{aligned}
\hat{U}(d)|\sigma\rangle_t &= \int_{t-\Delta}^{t+\Delta} dt' F(t'-t)(\alpha\, \hat{a}_{t'+d}^\dagger |g\rangle_a |g\rangle_b \\
&\quad + \beta\, \hat{b}_{t'+d}^\dagger |g\rangle_a |g\rangle_b) \\
&= |\sigma\rangle_{t+d} \quad (2)
\end{aligned}$$

Notice that in this representation qubit states at different clock-cycles are normalized over non-overlapping Hilbert spaces [11]. Thus tensor products of qubit states at different clock-cycles, such as $|\sigma\rangle_{t_{n-1}} \otimes |\sigma\rangle_{t_n}$ are mathematically allowed.

Using this formalism we can describe an entangled qubit pair at the $nth$ clock-cycle by:

$$|\phi^+\rangle_{t_n} = |0\rangle_{t_n,1}|0\rangle_{t_n,2} + |1\rangle_{t_n,1}|1\rangle_{t_n,2} \quad (3)$$

where we have introduced the additional subscripts, 1 and 2, to label different spatial locations of the two qubits. The details of the spatial wave-functions (assumed identical for the two qubits) are suppressed. For a free evolving entangled pair we could write this state as

$$\begin{aligned}
&... \otimes (|0\rangle_{t_{n-1},1}|0\rangle_{t_{n-1},2} + |1\rangle_{t_{n-1},1}|1\rangle_{t_{n-1},2}) \\
&\otimes (|0\rangle_{t_n,1}|0\rangle_{t_n,2} + |1\rangle_{t_n,1}|1\rangle_{t_n,2}) \\
&\otimes (|0\rangle_{t_{n+1},1}|0\rangle_{t_{n+1},2} + |1\rangle_{t_{n+1},1}|1\rangle_{t_{n+1},2}) \otimes ... \quad (4)
\end{aligned}$$

using the fact that we are allowed to take tensor products of different clock-cycles. What does this state mean? We consider projective measurements. If a measurement is made on both qubits at the $nth$ clock cycle then we are only interested in the part of the state describing that clock-cycle and should trace out the rest. This trace is of course trivial as there is no coupling between different clock-cycles, immediately giving us back Eq.3 upon which we can apply the standard projection formalism to obtain the probability of particular outcomes. These probabilities will show perfect correlation between the measurements on the two qubits. Now suppose a measurement was made on only qubit 1 at the $n$th clock-cycle giving the result 0. Projection onto the state $\langle 0|_{t_n,1}$ revises our knowledge of the state at, and downstream of the $n$th clock-cycle. Thus Eq.4 collapses to

$$... \otimes (|0\rangle_{t_{n-1},1}|0\rangle_{t_{n-1},2} + |1\rangle_{t_{n-1},1}|1\rangle_{t_{n-1},2})$$
$$\otimes |0\rangle_{t_n,2} \otimes |0\rangle_{t_{n+1},2} \otimes ... \quad (5)$$

and measurements of qubit 2 at subsequent clock-cycles will definitely give the result 0. Although an unusual description, this approach is equivalent to standard quantum mechanics within the confines of our simplified model.

## III. TIME DISPLACED ENTANGLEMENT

We now consider what this model predicts when one of the qubits suffers a relativistic time shift. Initially the two qubits are in the same inertial frame. Now suppose qubit 1 is taken on a round-trip at high velocity relative to this rest frame. After the trip the qubit is returned to the initial inertial frame. We expect from classical relativity theory that the time evolution of the qubit taken on the round-trip will have slowed relative to the stay at home qubit, thus effectively travelling into the future [12]. This can be modeled as a local application to qubit 1 of the time translation operator [13, 14]. For energy eigenstate qubits this effectively only produces a phase shift. However, for our broadband qubits it results in a shift in the local time of qubit 1. Suppose this time difference between the two qubits after the round trip, $\tau$, is equal to one clock-cycle time, ie: $\tau = t_n - t_{n-1}$. The entanglement will still be expressed as occurring between the $n$th clock-cycle of each qubit. However, due to the time dilation, the $(n-1)$th clock-cycle of qubit 1 now occurs at $t_n$ according to clocks in the initial inertial frame. We are thus led to write the following description of the entangled state after the round-trip:

$$|\bar{\phi}^+\rangle = |0\rangle_{t_n,1}|0\rangle_{(t_n-\tau),2} + |1\rangle_{t_n,1}|1\rangle_{(t_n-\tau),2}$$
$$= |0\rangle_{t_n,1}|0\rangle_{t_{n-1},2} + |1\rangle_{t_n,1}|1\rangle_{t_{n-1},2} \quad (6)$$

We refer to Eq.6 as time displaced entanglement and we will now study some unusual properties of this state. Both qubits are now again in the same inertial frame, so we can apply non-relativistic quantum techniques to their analysis.

In order to model measurements on the time displaced state we consider the tensor product of the state Eq.6 with itself after evolution $\hat{U}(\tau)$:

$$(|0\rangle_{t,1}|0\rangle_{t-\tau,2} + |1\rangle_{t,1}|1\rangle_{t-\tau,2})$$
$$\otimes (|0\rangle_{t+\tau,1}|0\rangle_{t,2} + |1\rangle_{t+\tau,1}|1\rangle_{t,2}) \quad (7)$$

where for simplicity we now just write $t$ for the $n$th clock cycle time. Again we note that this tensor product is allowed because the two states occupy non-overlapping Hilbert spaces. If a measurement is made on the system at clock time $t$, only those state components at time $t$ will contribute and components representing other times will be irretrievably lost. Thus we trace out the state components at other times. Unlike the example of the previous section the trace is now not trivial due to the coupling between different clock-cycles. We obtain the mixed state

$$\rho = 1/2(|0\rangle_{t,1}\langle 0|_{t,1} + |1\rangle_{t,1}\langle 0|_{t,1})$$
$$\otimes 1/2(|0\rangle_{t,2}\langle 0|_{t,2} + |1\rangle_{t,2}\langle 0|_{t,2}). \quad (8)$$

Projective measurements on this state will show no correlations. We thus predict that the entanglement will appear completely decohered as a result of the time displacement.

## IV. QUANTUM NON-LINEARITY

This result has quite unusual consequences when applied to a quantum circuit such as the one depicted in Fig.1. A qubit in an arbitrary state, $\alpha|0\rangle_{t',1} + \beta|1\rangle_{t',1}$ which for simplicity is taken to be a pure state, is entangled with another qubit, prepared in the zero state, using a Controlled-NOT (CNOT) gate. The resulting non-maximally entangled state is $\alpha|0\rangle_{t',1}|0\rangle_{t',2} + \beta|1\rangle_{t',1}|1\rangle_{t',2}$. Qubit 1 is subjected to a time dilating event leading to the state $\alpha|0\rangle_{t,1}|0\rangle_{t-\tau,2} + \beta|1\rangle_{t,1}|1\rangle_{t-\tau,2}$ at some later clock cycle $t$. A second CNOT is now performed between the two qubits. Proceeding as before we write a tensor product of the state and an evolved version such that we can apply the CNOT between state components with corresponding time signatures. The result after the CNOT is the state

$$\alpha^2|0\rangle_{t,1}|0\rangle_{t-\tau,2}|0\rangle_{t+\tau,1}|0\rangle_{t,2}$$
$$+ \alpha\beta|0\rangle_{t,1}|0\rangle_{t-\tau,2}|1\rangle_{t+\tau,1}|1\rangle_{t,2}$$
$$+ \beta\alpha|1\rangle_{t,1}|1\rangle_{t-\tau,2}|0\rangle_{t+\tau,1}|1\rangle_{t,2}$$
$$+ \beta^2|1\rangle_{t,1}|1\rangle_{t-\tau,2}|1\rangle_{t+\tau,1}|0\rangle_{t,2} \quad (9)$$

We now discard qubit 1. The state of qubit 2, at clock-cycle $t$, is then given by tracing out qubit 1 and the non-relevant time components of qubit 2. The solution for the output state of qubit 2 is then given by the density operator

$$\rho_2 = (|\alpha|^4 + |\beta|^4)|0\rangle_{t,2}\langle 0|_{t,2} + 2|\alpha|^2|\beta|^2|1\rangle_{t,2}\langle 1|_{t,2} \quad (10)$$

More generally if the input state is given by the arbitrary density operator $\rho_{in} = \gamma_{00}|0\rangle_{t',1}\langle 0|_{t',1} + \gamma_{11}|1\rangle_{t',1}\langle 1|_{t',1} + \gamma_{01}|0\rangle_{t',1}\langle 1|_{t',1} + \gamma_{10}|1\rangle_{t',1}\langle 0|_{t',1}$ then the output state will be

$$\rho_2 = (\gamma_{00}^2 + \gamma_{11}^2)|0\rangle_{t,2}\langle 0|_{t,2} + 2\gamma_{00}\gamma_{11}|1\rangle_{t,2}\langle 1|_{t,2} \quad (11)$$

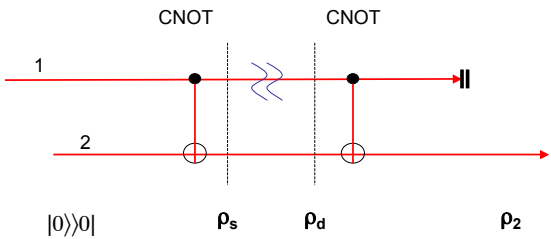
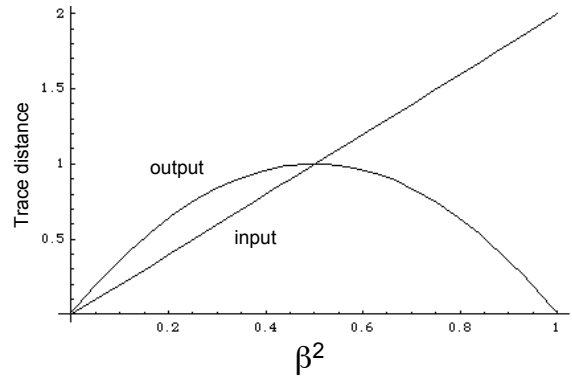

FIG. 1: Circuit for inducing non-linear evolution on an arbitrary qubit. The wavy lines indicate a period of relativistic time dilation.

FIG. 2: Plot of the trace distance between the logical zero state and an arbitrary real superposition, at the input and after the time-loop teleportation. For the region $\beta^2 < 0.5$ we have the extraordinary result that the trace distance is increased. Notice, though that for $\beta^2 > 0.5$ the trace distance is decreased. On average distinguishability is decreased.

Eq.11 (and Eq.10) features non-linear non-unitary evolution of the input qubit to the output qubit. Interestingly this is the same evolution as that predicted by Bacon [15], using the formalism of Deutsch [16] in the context of a CNOT + SWAP interaction between a free evolving qubit and a closed time-like curve generated by a quantum worm-hole.

The highly unusual properties of our system can be illustrated by considering how the trace distance $D = Tr|\rho_A - \rho_B|$ between the logical zero state and a real superposition state (i.e. $\alpha, \beta$ real) changes due to the evolution. At the input $D = 2\beta^2$. Suppose $D_a$ is the trace distance after evolution. Normally it is always true that $D_a \leq D$, meaning that the distinguishability between two quantum states cannot be increased. However from Eq.10 we have $D_a = 4(\beta^2 - \beta^4)$. As shown in Fig.2, distinguishability is increased in the region $0 < \beta^2 < 1/2$. Notice, though, that outside this region distinguishability is reduced. Indeed, on average, intergrating around the real great circle, distinguishability is reduced. Never-the-less, Bacon [15] has shown that significantly increased computing power is implied by evolution of this kind. With such an unusual evolution it is important to ask whether basic physical principles such as the second law of thermodynamics and relativistic "no-signalling" are being up-held.

## V. ENTROPY AND NO-SIGNALING

We consider first how the entropy of the system changes under this evolution. First notice that the state Eq.9, although quite unusual, is a pure state and thus has zero entropy, just like the input state. This indicates that the evolution is in fact unitary and that it could be reversed (from this point) by applying the appropriate inverse. Explictly this inverse would involve another time dilation, applied to qubit 2, followed by a CNOT with again qubit 1 as the control.

Secondly let us consider the situation when we assume we have lost access to all components not at time $t$. That is we will not allow any further time dilations to be applied before measurement. Consider the state after the first CNOT ( but before the time dilation), for some arbitrary input state. Suppose this (possibly entangled) state is described by $\rho_s$ with entropy $S(\rho_s)$ (see Fig.1). After the time dilation, and tracing out inaccessible time components, it can be shown quite generally that the resultant density operator is $\rho_d = Tr_2[\rho_s] \otimes Tr_1[\rho_s]$ with entropy $S(\rho_d) = S(Tr_2[\rho_s]) + S(Tr_1[\rho_s])$. But subadditivity requires that $S(\rho_s) \leq S(Tr_2[\rho_s]) + S(Tr_1[\rho_s])$ [4] and hence $S(\rho_s) \leq S(\rho_d)$. CNOTs are unitary and so do not change the entropy, thus we can conclude quite generally that in going from the initial to final states (prior to any measurements) the entropy of the system does not decrease.

Finally we consider the complete evolution from input qubit to output qubit described by Eq.10. In this case examples can be found in which the entropy of the qubit decreases, as shown in Fig.3. However, we have now included an effective generalized measurement on one of the qubits (the tracing out of qubit 1). It is a standard result that entropy may decrease for generalized measurements [4] , with the waste entropy accumulated by the measurement device. In other words, any decrease in entropy that occurs in evolving from $\rho_d$ to $\rho_2$ is just that expected from standard quantum mechanics.

We now turn to the question of whether the non-linear evolution can lead to faster than light signaling. At first sight it may seem this is possible via the following argument: Suppose Alice and Bob, who are far appart, share an ensemble of entangled states of the form $|00\rangle + |11\rangle$. Alice either measures her ensemble of states in the computational basis, thus non-locally preparing Bob's states as $|0\rangle$ or $|1\rangle$, or she measures her qubits in the diagonal basis, thus preparing Bob's qubits into the states $|0\rangle + |1\rangle$ or $|0\rangle - |1\rangle$. Bob then sends his qubits through the circuit of Fig.1. According to Eq.10, if Alice measures in the computational basis Bob's qubits will alswayss have the value 0, but if Alice measures in the diagonal basis Bob will get a qubit value of 1 in 50% of the cases. It seems that by analysing the statistics of the results Bob could learn what Alice did in a superluminal way. However, this is incorrect as we now show.

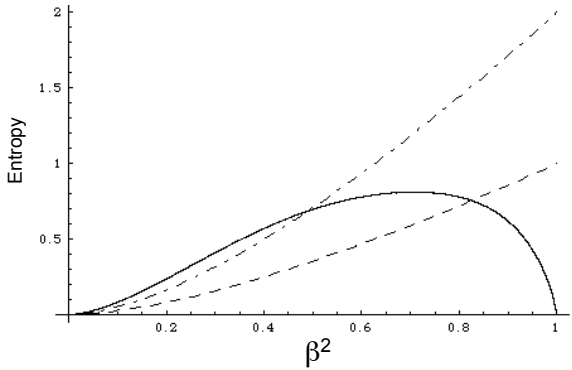

FIG. 3: Examples of the entropic considerations discussed in the text. Suppose the input state is a classical mixture of the vacuum state and the state $\alpha|0\rangle_{t,1} + \beta|1\rangle_{t,1}$. The entropy of this input state is plotted as a function of $\beta^2$ as the dashed curve. If all time components are considered and no qubits are traced out the entropy remains the same after the time dilation. If the inaccessible (without time dilation) time components are traced out we obtain the dot-dashed curve for $\rho_d$ (see Fig.1) and entropy increases. If we consider the entire evolution including the trace out of qubit 1 we obtain the solid curve. The change in entropy that occurs in going from the dot-dashed curve to the solid one is just that expected for standard quantum evolution.

The state shared by Alice and Bob should be written

$$(|0\rangle_{t-\tau,a}|0\rangle_{t-\tau,b} + |1\rangle_{t-\tau,a}|1\rangle_{t-\tau,b}) \\ \otimes (|0\rangle_{t,a}|0\rangle_{t,b} + |1\rangle_{t,a}|1\rangle_{t,b}) \quad (12)$$

where we are anticipating the need for information about the other time components in the analysis. Suppose Alice obtains the measurement result $\gamma|0\rangle + \delta|1\rangle$, where $\gamma$ and $\delta$ can be choosen to coincide with any of the results obtained from the two bases. Bob's state becomes

$$(|0\rangle_{t-\tau,a}|0\rangle_{t-\tau,b} + |1\rangle_{t-\tau,a}|1\rangle_{t-\tau,b}) \\ \otimes (\gamma|0\rangle_{t,b} + \delta|1\rangle_{t,b}) \quad (13)$$

Notice that Alice's measurement does not collapse the entanglement at the earlier clock-cycle. Sending this state through the circuit of Fig.1 results in an output state for Bob's qubit of

$$\rho_b = 1/2(|0\rangle\langle 0| + |1\rangle\langle 1|) \quad (14)$$

which conveys no information to Bob about Alice's measurement basis. This is also the result we would get if we took Bob's input state to be the reduced density operator of their shared entangled state. That is, if we took Bob's initial state to be

$$\rho_b = \rho_{rb,t-\tau} \otimes \rho_{rb,t} \quad (15)$$

where $\rho_{rb}$ is the reduced density operator of Bob and Alice's state, then equivalent results would be obtained. Thus our model rules out superluminal signaling and retains the standard equivalence between the collapse and reduced density operator formalism. On the other hand notice that a measurable difference between proper and improper mixtures is predicted. In contrast to the non-locally produced state of Eq.13, a locally produced state can be represented by

$$(\gamma|0\rangle_{t-\tau,b} + \delta|1\rangle_{t-\tau,b}) \otimes (\gamma|0\rangle_{t,b} + \delta|1\rangle_{t,b}) \quad (16)$$

and this representation does not change dependent on whether the value of the classical parameters ($\gamma$ and $\delta$) used to produce the state are known or unknown. Thus, for example, a classical mixture of the locally produced states $|\phi\rangle$ and $|\psi\rangle$ should be represented

$$\rho_l = P_\phi(|\phi\rangle_{t-\tau} \otimes |\phi\rangle_t \ h.c.) + P_\psi(|\psi\rangle_{t-\tau} \otimes |\psi\rangle_t \ h.c.) \quad (17)$$

where $h.c.$ stands for Hermitian conjugate and the $P_i$ are the classical probabilities of producing each state. Eqs 15 and 17 are distinct and will in general give different answers in situations in which standard quantum mechanics would hold them equivalent.

## VI. CONCLUSION

We have described an interaction which combines quantum entanglement with relativistic time dilation effects. We have shown that the resulting evolution lies outside the realm of standard quantum mechanics. In particular non-linear state evolution is predicted. However, the general class of evolutions possible from this effect do not lead to violations of causality or the principles of thermodynamics.

The analysis presented here is a toy model in both its physical abstraction and in the unphysical assumption of a strict bounding of the qubit time spectra (see note [9]). This assumption can be relaxed by adopting a more sophisticated analysis based on expectation values and Gaussian temporal wave functions. The current results then emerge in the limit of time dilations large in comparison to the widths of the Gaussians. This approach will be detailed elsewhere. Another key assumption leading to Eq.6 is that after the round-trip, with both qubits again in the same inertial frame, the only effect on the travelling qubit is the classically predicted time dilation. Although this proposal would obviously be demanding to test experimentally it does not seem beyond the realm of horizon technology, with maintenance of coherence during the time dilation the likely biggest hurdle.

## Acknowledgments

I thank P.P.Rohde, C.M.Savage, K.Pregnell, A.Lund, T.Downes, D.Gottesman and G.J.Milburn for helpful discussions. This work was supported by the Australian Research Council.